# Words or Numbers? How Framing Uncertainties Affects Risk Assessment and Decision-Making

*Robin Bodenberger[1] & Kirsten Thommes[1]*

[1]Chair of Organizational Behavior

Paderborn University

Warburger Straße 100

33100 Paderborn

Germany

robin.bodenberger@uni-paderborn.de (R. Bodenberger)

https://orcid.org/0009-0000-6110-870X

kirsten.thommes@uni-paderborn.de (K. Thommes)

https://orcid.org/0000-0002-8057-7162

Corresponding author: Robin Bodenberger

# Words or Numbers? How Framing Uncertainties Affects Risk Assessment and Decision-Making

Abstract

Senders of messages prefer to communicate uncertainty verbally (e.g., something is likely to happen) rather than numerically (such as 75%), leaving receivers with imprecise information. While it is well established that receivers translate verbal probabilities into numerical values that systematically deviate from the intended numerical meaning, it is less clear how this discrepancy influences subsequent behavioral actions. Thus, the role of verbal versus numerical communication of uncertainty warrants additional attention, to investigate two critical questions: 1) whether differences in decision-making under uncertainty arise between these communication forms, and 2) whether such differences persist even when verbal phrases are translated accurately into the intended numerical meaning. By implementing a laboratory experiment, we show that individuals place significantly lower values on uncertain options with medium to high likelihoods when uncertainty is communicated verbally rather than numerically. This effect may lead to less rational decisions under verbal communication, particularly at high likelihoods. Those results remain consistent even if individuals translate verbal uncertainty correctly into the intended numerical uncertainty, implying that a biased behavioral response is not only induced by miscommunication. Instead, ambiguity about the exact meaning of a verbal phrase interferes with decision-making even beyond potential mistranslations. These findings tie in with previous research on ambiguity aversion, which has predominantly operationalized ambiguity through numerical ranges rather than verbal phrases. Based on our findings we conclude that managers should communicate uncertainty numerically, as verbal communication can unintentionally influence the decision-making process of employees.

## 1. INTRODUCTION

In times of global crises, institutional upheaval and dynamic environments, managers and policymakers need to openly talk about uncertainties to achieve trustworthy communication. One critical aspect of this process is choosing how to express uncertainty, which can be done in two ways: either via verbal phrasing ('likely') or via numbers ('there is a 75% likelihood'). Even though past research has frequently highlighted that the communication of uncertainty is tricky (see for a review Spiegelhalter, 2017), research on the effects of communication practices on subsequent decision-making is still in its infancy. Particularly, the behavioral consequences of expressing uncertainty as a number versus a verbal phrase are not fully understood yet.

Communication inherently involves two parties: the sender and a receiver. Notably, they differ in their preferences on how to communicate uncertainty. Senders of messages prefer to use verbal phrases over numbers when conveying uncertainties (Erev & Cohen, 1990; Wallsten et al., 1993; Juanchich & Sirota, 2020a, 2020b; Rosen et al., 2021), as the vagueness of verbal expressions enables them to maintain credibility after erroneous forecasts (Juanchich et al., 2012; Dhami & Mandel, 2022). In contrast, receivers favor numerical expressions of uncertainty (see also Brun & Teigen, 1988; Wallsten et al., 1993; Andreadis et al., 2021), as they appreciate the precision numbers can provide (Lofstedt et al., 2021). This mismatch between sender and receiver is known as the communication mode preference paradox (see Erev & Cohen, 1990).

The Intergovernmental Panel on Climate Change (IPCC) provides a prominent example of a sender opting for verbal probabilities. A guidance note published in 2010 instructed IPCC authors to communicate quantifiable uncertainties via verbal probabilities rather than precise numerical values (Mastrandea et al., 2010), establishing this as an intentional practice which is reflected in the most recent reports published in 2014 (IPCC, 2014) and 2023 (IPCC, 2023) respectively. This communication choice motivates our research as it creates practical challenges for receivers who need to adapt to a format, they neither prefer nor process easily (e.g., see Budescu et al., 2012; 2014; Kause et al., 2022).

Particularly, receivers face difficulties dealing with verbal probabilities, as these lack standardized definitions and are open to individual interpretation. Past research has consistently emphasized the ambiguity of such phrases, revealing that receivers systematically misinterpret them in ways unintended by the sender. These discrepancies have been documented across various domains (see Teigen et al., 2022), including meteorology (Rosen et al., 2021), the medical field (Brun and Teigen; 1988; Berry et al., 2002, Andreadis et al., 2021), and reports on climate change (Budescu et al., 2012; 2014).

Many previous studies have focused on reducing misunderstandings caused by verbal uncertainty communication. Budescu et al. (2012; 2014), for example, show that supplementing verbal phrases with numerical ranges can improve congruence between sender and receiver. While these efforts help clarify interpretation, they leave open a more fundamental question: are misunderstandings reflected in subsequent behavioral actions? And if so, does reducing them lead to different decisions?

In this paper we explore how the misalignment between verbal and numerical probabilities leads to behavioral consequences in subsequent decision-making. We aim to understand whether potential differences in decision-making arise due to humans' difficulty translating verbal probabilities into the intended numerical meaning, or whether potential behavioral biases are driven by the verbal communication form itself and persist even when translations are accurate. We evaluate the effects of verbal and numerical uncertainty communication by using a laboratory experimental setting, analyzing individuals' actions after receiving either numerical or verbal probabilities.

First, our findings confirm that verbal probabilities are perceived differently among receivers, often deviating from the intended numerical meaning. Second, differences in decision-making under uncertainty arise between verbal and numerical communication for medium and high likelihoods. Specifically, individuals assign lower reservation prices when uncertainty is communicated verbally rather than numerically, leading to less rational decisions at high likelihoods with evaluations diverging from the expected value of the uncertain option. Third, this holds true even if individuals translate verbal probabilities into the intended numerical value. The latter finding suggests that responses in behavior may not be cured by educational means but are inherent to verbal communication.

## 2. RELATED LITERATURE

To understand how verbal and numerical communication of uncertainty shape decision-making, it is essential to examine the distinct psychological characteristics of these communication formats. While numerical probabilities offer a standardized and widely interpretable format, their verbal counterparts are inherently more complex as they have no fixed numerical meaning and are open to subjective interpretation. This interpretative uncertainty introduces ambiguity – a key characteristic of verbal probabilities.

Ambiguity is a well-established concept in the context of decision-making under uncertainty. Ellsberg (1961) first introduced the terminology by differentiating between risky options with known probabilities, and ambiguous options that additionally impose uncertainty regarding the exact probability value. In his experiment, participants chose between two urns, each containing 100 balls in either black or red. The first urn, representing the risky option, contained an equal distribution of 50 balls per color. The second, ambiguous urn had an unknown distribution, with the number of black and red balls ranging anywhere from 0 to 100. Participants consistently preferred to bet on the risky urn, avoiding the ambiguous one. Ellsberg (1961) termed this behavior as ambiguity aversion, a cognitive bias where individuals prefer known risks over uncertain ones, even when the expected outcomes are identical. In fact, ambiguity aversion can promote irrational behavior leading decision-makers to not choose the option with the highest expected value (Ellsberg, 1961).

This conceptual separation between risk and ambiguity has motivated a growing body of research that builds on Ellsberg's foundational work. A meta study by Bühren et al. (2023) showcases the growing interest in ambiguity aversion with publications on the topic increasing rapidly in the 2000's. In this timeframe many applications and extensions of Ellsberg's design were published. Halevy (2007) and Borghans et al. (2009) implement multiple urns with a consistent expected payoff but varying degrees of ambiguity across urns. Both studies expand on Ellsberg's approach by moving beyond a binary choice between risky and ambiguous urn and instead derive ambiguity preferences via a Becker-DeGroot-Marschak (1964) mechanism. Hereby, participants are tasked to select a reservation price for each urn indicating a guaranteed price at which they are indifferent between a certain payout and the uncertain payout of the risky/ambiguous urn. Their results confirm Ellsberg's findings as participants select lower reservation prices for ambiguous urns compared to a risky urn. More specifically, individuals select lower reservation prices as urns become more ambiguous with wider probability ranges (Borghans et al., 2009). Thereby, leading to less rational decisions, with reservation prices deviating further from the expected value of the lottery (Borghans et al., 2009). Furthermore, Kocher et al. (2018) considered ambiguity aversion more extensively and expand Ellsberg's framework by considering the possibility of both gains and losses as well as different probability levels. Their results show that ambiguity aversion is not universal and may only hold in the domain of gains for moderate to high likelihoods (Kocher et al., 2018).

Focusing on Ellsberg's (1961) original design as well as its various extensions, ambiguity is exclusively operationalized through missing probabilities or numerical ranges. Verbal probabilities – though inherently ambiguous – have not been considered in this line of work.

This represents both a conceptual gap in the literature on ambiguity aversion as well as a missed opportunity to better understand decision-making under uncertainty through a lens of real-world uncertainty communication via verbal phrases.

Ambiguity attitudes and risk attitudes are highly correlated (e.g., see Buckholtz et al., 2017; Wu et al., 2024), yet they are conceptually distinct. Wu et al. (2024), for example, show that an intervention designed to reduce compound risk aversion had no effect on ambiguity aversion, underscoring that the two phenomena are psychologically separable. Complementary, previous research has consistently shown that attitudes towards both risk and ambiguity affect choice behavior (e.g., see Ghosh & Ray, 1992; 1997). While some recent research has started to emphasize the relation of verbal probabilities and risk attitudes (e.g., see Milczarski et al., 2025), ambiguity attitudes remain unexplored but may provide a more fitting theoretical framework.

Given the conceptual and psychological distinction between risk and ambiguity, and the inherent ambiguity of verbal probability phrases, we adopt ambiguity aversion as our theoretical framework to explain how the two formats of uncertainty communication influence decision-making. Based on previous research on ambiguity aversion we propose that receivers are averse to verbal probabilities as they are ambiguous and derive the following hypotheses:

> *H1: Receivers assign lower values to uncertain options when their likelihood is communicated verbally rather than numerically*
>
> *H2: Receivers exhibit less rational behavior when uncertainty is communicated verbally rather than numerically*

Numerical probabilities represent a clear and precise foundation for decision-making, whereas verbal probabilities are prone to mistranslations, potentially creating a biased basis for subsequent decisions. However, even when verbal probabilities are perceived correctly and translated into the intended value, ambiguity remains for the receiver as they do not know whether they correctly translated the phrase or not. Thus, we further assume that differences in decision-making between verbal and numerical communication remain even when verbal probabilities are translated in accordance with the intended numerical probability, and hypothesize:

> *H3: Receivers assign lower values to uncertain options when their likelihood is communicated verbally rather than numerically even when the verbal probability is translated correctly*

> *H4: Receivers exhibit less rational behavior when uncertainty is communicated verbally rather than numerically even when the verbal probability is translated correctly*

## 3. METHOD

We conducted an experiment designed to study the impact of numerical and verbal probability phrases on subsequent decision-making. The framework of the experiment is adapted from an approach developed by Halevy (2007) based on Ellsberg's (1961) original research on ambiguity aversion. Using this approach, we can derive the value an individual is willing to place on risky or ambiguous options. Our experiment deviates from Halevy's (2007) design by implementing ambiguity through verbal probability phrases rather than numerical ranges.

### 3.1. Setting

We conducted a randomized experiment followed by a questionnaire, both implemented as an online study via the platform SoSci Survey. The experiment received ethical approval from the ethics committee of Paderborn University, and all subjects provided informed consent to participate in the experiment. If subjects did not consent to participate, they could not start the experiment.

Participants were recruited through Prolific and limited to individuals from the UK to ensure native English speakers and avoid potential language barriers for verbal uncertainty communication. Additionally, we implemented two comprehension checks in accordance with Prolific guidelines to ensure that participants understood the procedure. Participants who failed a comprehension check more than once were asked to stop the study and return their submission. In total, 229 individuals started the study, out of which 29 ended their participation due to failed comprehension checks, while the remaining 200 participants completed the study taking about 10 minutes on average. Among the participants 54% were male (46% female). The average age was 35.67 years, ranging between 18 and 75. For fully completing the study, participants were paid with a combination of fixed and variable payments and earned £3.08 on average.

### 3.2. Communication of Uncertainty

In the experiment, participants are randomly assigned to receive uncertainty information either in a verbal or a numerical format. The verbal group receives verbal probability phrases used by the IPCC in their official reports (see IPCC, 2014; 2023). To avoid overburdening participants – who complete a decision task for each phrase – we do not adopt all ten phrases historically used by the IPCC. Instead, we focus on a subset of five that span a wide range of likelihood levels.

The numerical group receives corresponding precise numerical probabilities which fall within the intervals of the interpretation guidelines provided by the IPCC as well as the range of mean numerical translations across different studies (e.g., Theil, 2002; Barnes 2016). To maintain consistency, we choose values that are approximately evenly distributed across the range of 1% to 99%. Specifically, we consider probabilities for the five likelihood levels displayed in Table 1.

Table 1. Verbal and Numerical Communication

| Likelihood level | Numerical probability | Verbal Probability | IPPC Interpretation Guideline |
|---|---|---|---|
| Very low | 1% | Exceptionally unlikely | 0-1% |
| Low | 25% | Unlikely | 0-33% |
| Medium | 50% | About as likely as not | 33-66% |
| High | 75% | Likely | 66-100% |
| Very high | 99% | Virtually certain | 99-100% |

Notably, choosing a corresponding value for the numerical group that ensures comparability between the two groups is considered to be a key challenge. Verbal probability estimates vary notably across individuals, not just within but also between studies (see Theil, 2002, for a meta study). Thus, adopting numerical translations from previous studies may not be sufficient. Mandel et al. (2021) address this issue by conducting the verbal condition first and then using the median of participants' numerical translations in their own sample to inform the numerical group. While this approach enhances previous study design by not relying on external estimates, it is still based on the assumption that a single numerical translation – such as the median – adequately represents all participants in the verbal group.

Meanwhile, participants in the verbal group are highly heterogeneous regarding their numerical translations. Even if the median value ensures comparability between groups on average, it inevitably misrepresents those participants whose translations deviate substantially from the median. Realistically, the two groups can at best be comparable across a subset of participants. As a result, we consider additional methods after data collection. Specifically, we implement a matching procedure focusing our analysis on participants in the verbal group whose self-reported interpretations align with the numerical values provided in the control group.

### 3.3. Experimental design

The experimental design expands on Halevy's (2007) take on ambiguity aversion. Halevy implemented urns with ten balls in two different colors, allowing for probabilities of drawing a specific color to range between 0 and 100% in intervals of 10%. Extending this setup to accommodate our full range of likelihood levels (e.g., 99%) would require an impractically large number of balls. To preserve the probabilistic structure while maintaining feasibility, we instead introduce uncertainty through urns containing ten lottery tickets.

The experiment then consists of five rounds. In each round, participants are confronted with a different urn and tasked to select one lottery ticket, which can either be a 'Win' (corresponding to a potential bonus payment of 200 pence) or a 'Loss' (0 pence). The probability of drawing a winning ticket varies across rounds and corresponds to one of five likelihood levels displayed in Table 1. Participants encounter each likelihood level once, in a randomized order.

Each round follows a two-stage procedure. In the first stage (ticket selection), participants are shown the urn alongside information about the likelihood of drawing a winning ticket. This likelihood is communicated either verbally (e.g., likely) or numerically (e.g., 75%), depending on the participant's randomly assigned treatment group. Afterwards, participants get to interact with the urn and choose one of the ten lottery tickets. Notably, the outcome of the ticket is not revealed immediately.

In the second stage (valuation task), a follow-up task is implemented to elicit the preference of participants. Specifically, we follow Halevy's (2007) approach by implementing a Becker-DeGroot-Marschak (1964) mechanism to identify at which guaranteed price participants are indifferent between a certain payout and the uncertain payout of their risky/ambiguous ticket (i.e., 200 pence for a 'Win' and 0 pence for a 'Loss'). Within this mechanism participants must select a reservation price indicating the lowest price at which they would be willing to sell their ticket. This price is then compared to a randomly generated offer ranging between 0 and 200 pence. If the offer exceeds the reservation price, the lottery ticket is sold, and participants are guaranteed to receive the offered money. Otherwise, the ticket is played out and the bonus payment is determined by the uncertain outcome of the lottery.

Note that one round of the experiment is randomly selected for bonus payment at the end of the experiment. Only then do participants learn which round was chosen for payment, whether their ticket was a 'Win' or a 'Loss', and whether it was sold or played. The experimental procedure is visualized in Appendix A.

After participants completed five rounds of drawing a ticket and listing their reservation price, those participants in the verbal communication group are additionally tasked to indicate their numerical translation for the five verbal probability phrases used in the experiment.

### 3.4. Description of variables

The dependent variables in our experiment are the reservation prices that participants selected for the different likelihood levels. The reservation price reveals information about the preference of the participants regarding the risky/ambiguous lottery ticket beyond a simple binary choice between risky and ambiguous options (Halevy, 2007). Additionally, we observe the rationality of the reservation price by considering its deviation to the expected value of the lottery.

The main independent variable within our experiment is the communication form of uncertainty, which is based on a randomly assigned treatment. In the verbal group we additionally tasked participants to indicate numerical translations for each phrase, to match participants between verbal and numerical group.

Additionally, we consider relevant covariates. Here, we include the participant's gender as previous research has shown that women are more risk averse and ambiguity averse than men (see Borghans et al., 2009). Moreover, fluid intelligence and cognitive abilities are commonly linked with decision-making under uncertainty (see Lilleholt, 2019). We account for fluid intelligence through the Berlin Numeracy Task (BNT) as first implemented by Cokely et al. (2012). BNT depicts an individual's ability to understand and evaluate uncertainties and is therefore especially relevant in the context of decision-making under uncertainty. Here, Cokely et al. (2018) emphasized its strong predictive power and efficiency, as it is based on an adaptive design with 2-3 questions yielding a score between 1 and 4 as shown in Appendix B. Moreover, we consider age as indicator of crystallized intelligence as proposed by Krefeld-Schwalb et al. (2024) to account for an age-related decline in fluid intelligence. Furthermore, we account for attentiveness through the logarithmic response time of a given task as Krefeld-Schwalb et al. (2024) argue that participants with low attentiveness may speed through a survey or experiment and miss crucial information, affecting their decision-making process.

### 3.5. Coarsened Exact Matching

Individuals vary considerably in how they translate verbal probabilities, leading to many translations that deviate from the intended numerical value. This variation poses a challenge for isolating the effect of the communication format itself, as any observed differences in decision-making between groups may arise not only from the communication form, but also from misalignment between intended and perceived probability in the verbal group. To improve comparability between groups and better identify the causal effect of the communication form, we apply a matching procedure.

Specifically, we implement coarsened exact matching (CEM) to create more comparable groups by matching participants in the verbal and numerical groups on relevant covariates. Crucially, we do not match solely on translation accuracy but also on additional factors that may affect decision-making. Specifically, we include age as it has a significant effect on numerical translation of various verbal phrases in our experiment (see Appendix C) and gender, which has been shown to affect decision-making under uncertainty in previous research (e.g., see Borghans, 2009) and is unequally distributed between the two groups in our sample.

CEM is particularly well-suited for this setting, as it overcomes the limitations of exact matching with (multiple) continuous variables where finding exact matching partners becomes unlikely due to the high number of unique combinations. Rather than requiring identical values across covariates, CEM temporarily coarsens each variable into intervals. Observations are then grouped into strata based on unique combinations of these coarsened values. Within each stratum, observations from verbal and numerical group are matched and subsequently assigned weights based on the number of observations of each group (see Iacus et al., 2012). Strata that contain units from just a single communication form are given a weight of zero (see Blackwell et al., 2009). While CEM itself is purely a data-processing method, the resulting weights are applied in our analysis. A visual overview of the matching process for each phrase is provided in Appendix D.

This approach offers distinct advantages over an exact matching method that is solely based on translation accuracy. First, by matching not only on translation accuracy but also on relevant covariates, CEM improves group comparability on multiple dimensions. Second, observations with only minor translation inaccuracies can still be matched and receive a corresponding weight, reducing data loss.

Notably, we expect a high degree of variation in how verbal probabilities are translated into numerical values, often going beyond minor inaccuracies. Thus, despite the advantages of CEM, we still expect some data loss. To account for this, we assigned a larger number of participants to the verbal group (149 participants) compared to the numerical group (51 participants).

## 4. RESULTS

In the following we will discuss the results of our experiment. We first focus on the translation of verbal probability phrases. Afterwards, we turn to decision-making and how it is impacted differently by verbal and numerical communication of uncertainty, and whether such differences remain after adjusting for translation accuracy.

Figure 1 illustrates density functions for the numerical translations of verbal phrases as indicated in the verbal communication group. For three phrases (unlikely, about as likely as not, and likely) the peak of the density function corresponds to the intended numerical meaning that is given to the control group while the remaining two phrases (exceptionally unlikely and virtually certain) are not more than 3 percentage points off. Despite this general fit, the verbal and numerical group cannot be considered comparable across all participants. Instead, all functions display notable variation around the peak, highlighting heterogeneity in translations and ambiguity of verbal probabilities.

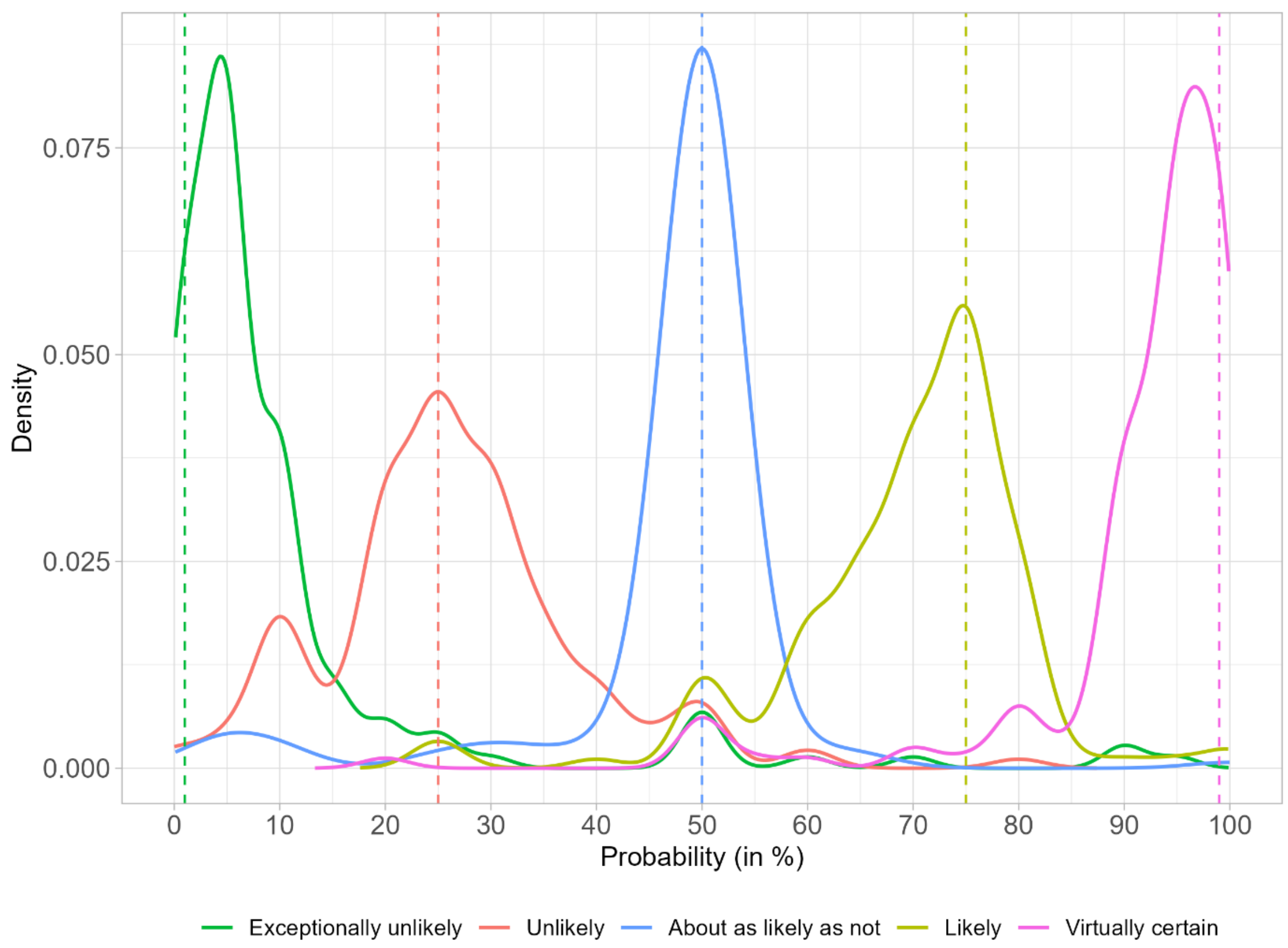

Figure 1. Density functions for the translations of verbal probability phrases

We account for heterogeneity in numerical translations by applying the matching procedure CEM. Table 2 depicts descriptive statistics of the numerical translations before and after the matching procedure is applied. The unmatched data reveals substantial variability in the translation of verbal probabilities, with standard deviations ranging between 11.41 and 16.55 percentage points. Additionally, all five phrases exhibit outliers on both ends of the scale, reflected in the wide range of responses. We mitigate this heterogeneity by applying CEM and matching on the intended numerical translation as well as age and gender as additional covariates. This approach substantially reduces variability: on average across all five phrases the average standard deviation is reduced from 12.77 to 2.37 percentage points, and the range narrows from 85.6 to 8.26 percentage points. Overall, the matching procedure leads to greater comparability between the verbal and numerical groups. These improvements come at the expense of data loss, with an average of 57 observations per phrase being removed from the verbal group.

Table 2. Descriptive statistics – Translation of verbal phrases

| Verbal phrases | Matching | Median | Mean | Std. dev. | Min | Max | Range | N |
|---|---|---|---|---|---|---|---|---|
| **Exceptionally Unlikely (1%)** | Before | 5 | 10.31 | 16.55 | 0 | 95 | 95 | 149 |
| | After | 5 | 4.92 | 3.26 | 0 | 10 | 10 | 118 |
| **Unlikely (25%)** | Before | 25 | 26.51 | 12.08 | 0 | 80 | 80 | 149 |
| | After | 25 | 27.10 | 2.49 | 25 | 30 | 5 | 61 |
| **About as likely as not (50%)** | Before | 50 | 47.31 | 11.70 | 2 | 100 | 98 | 149 |
| | After | 50 | 49.68 | 1.74 | 40 | 50 | 10 | 124 |
| **Likely (75%)** | Before | 70 | 69.46 | 11.42 | 25 | 100 | 75 | 149 |
| | After | 75 | 76.38 | 2.25 | 75 | 80 | 5 | 61 |
| **Virtually certain (99%)** | Before | 95 | 91.67 | 12.09 | 20 | 100 | 80 | 149 |
| | After | 98 | 97.10 | 2.13 | 92 | 100 | 8 | 97 |
| **Average** | Before | | | 12.77 | | | 85.60 | 149 |
| | After | | | 2.37 | | | 8.26 | 92.2 |

Intended numerical probabilities used in the numerical group are displayed in parentheses

In the following, we shift the focus from participants' translation to their subsequent decision-making. In the experiment, participants were tasked to select reservation prices for lottery tickets, based on communicated likelihoods of winning. Figure 2 provides a descriptive overview of these reservation prices for both numerical and verbal communication, including separate bars for the verbal condition depending on translation accuracy (before and after matching). Across both communication modes, participants responded systematically to variations in the communicated likelihood: higher probabilities of winning (e.g., 'likely' vs. 'about as likely as not') led to higher reservation prices. Pairwise comparisons of reservation prices between any two likelihood levels reveal statistically significant differences, with p-values smaller than 0.01 according to a Wilcoxon signed rank sum test. Though, when increasing the likelihood of a 'Win', the average reservation price does not increase at the same rate as the expected value of the lottery. Instead, individuals deviate from the expected value and rational response by acting risk-seeking at low likelihoods and becoming increasingly risk-averse at higher likelihood levels.

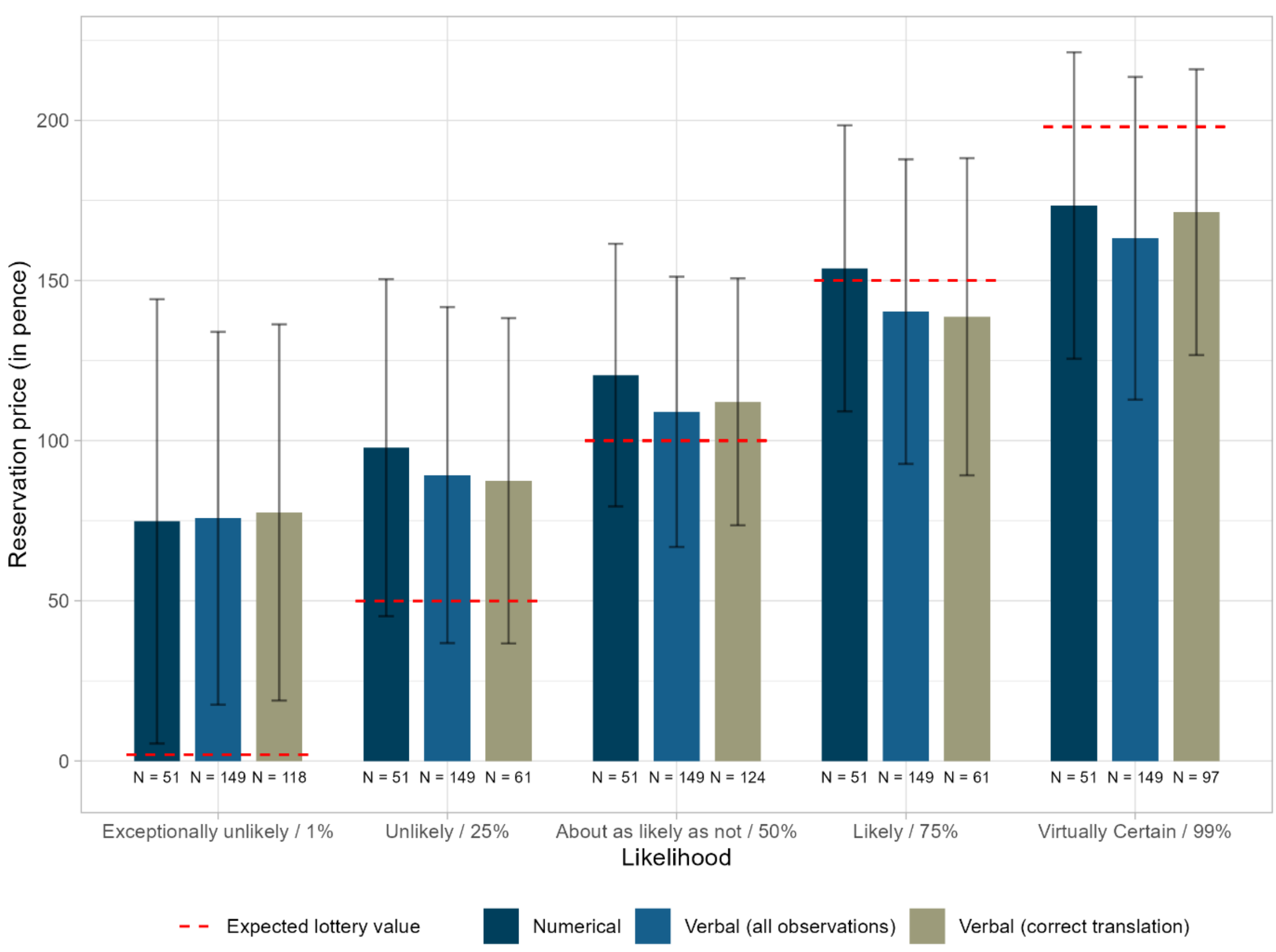

Figure 2. Average reservation prices based on numerical and verbal communication

Notably, decisions may vary not only between likelihood levels but also between communication forms at a fixed likelihood level. To examine these differences, we first apply a Mann-Whitney U test to compare the distribution of reservation prices between the numerical and verbal communication groups. Because this non-parametric test does not support weighting, we restrict the analysis to the unmatched data, as the matched dataset incorporates weights from the CEM procedure. Based on the unmatched sample, we find statistically significant differences between the verbal and numerical groups at a medium, high and very high likelihood at a 10% level as shown in Table 3.

Table 3. Mann-Whitney U test for differences in reservation prices

| Likelihood* | Expected Value | Mean Reservation Price | | p-Value |
|---|---|---|---|---|
| | | Numerical (N=51) | Verbal (N=149) | |
| **Exceptionally Unlikely (1%)** | 2 | 74.784 | 75.779 | 0.495 |
| **Unlikely (25%)** | 50 | 97.824 | 89.255 | 0.284 |
| **About as likely as not (50%)** | 100 | 120.471 | 108.993 | 0.077 |
| **Likely (75%)** | 150 | 153.784 | 140.295 | 0.075 |
| **Virtually certain (99%)** | 198 | 173.412 | 163.161 | 0.092 |

*The numerical probabilities proposed to the control group are displayed in parentheses

To further investigate these differences, we turn to regression models as a parametric approach, which allows us to remove decisions based on mistranslations by incorporating the weights retrieved from CEM. Specifically, we run two regression models for each level of likelihood – one with weights applied (after matching) to capture the differences between verbal and numerical communication when verbal phrases are translated as intended, complemented by models with all observations (before matching) to assess the robustness of the effect. This approach enables us to examine the impact of verbal versus numerical communication on decision-making and assess the importance of translation accuracy.

Table 4 displays the results of a regression model for each of the likelihood levels after the matching process with CEM weights applied. The results of the models reveal that on average and c.p. individuals select significantly lower reservation prices for medium and high likelihoods when uncertainty is communicated verbally rather than numerically. Specifically, participants in the verbal condition who translate the phrases ‘about as likely as not’ and ‘likely’ close to their intended probabilities of 50% and 75%, set lower reservation prices by 13.949 and 23.482 pence respectively, compared to participants who received these probabilities in numerical form.

Table 4. Regression models for reservation prices (after matching)

| Explanatory variables | (1) Very low likelihood | (2) Low likelihood | (3) Medium likelihood | (4) High likelihood | (5) Very high likelihood |
|---|---|---|---|---|---|
| Verbal | -3.445 | -10.375 | -13.949** | -23.482** | -6.353 |
| (0=numerical) | (11.179) | (11.174) | (7.017) | (10.445) | (9.928) |
| Male | -4.785 | -5.975 | -1.838 | -8.693 | -14.982 |
| (0=female) | (10.392) | (11.121) | (7.003) | (10.932) | (10.000) |
| Age | -0.235 | 1.328** | -0.224 | -0.251 | -0.546 |
| | (0.478) | (0.612) | (0.338) | (0.482) | (0.459) |
| BNT (1st quartile – ref.) | | | | | |
| 2nd quartile | -20.214 | -13.079 | -8.003 | -4.439 | 2.373 |
| | (15.822) | (15.149) | (10.229) | (16.201) | (16.784) |
| 3rd quartile | -61.971*** | -29.233* | -13.583 | -15.826 | 14.382 |
| | (16.344) | (17.088) | (11.866) | (16.071) | (13.136) |
| 4th quartile | -38.972*** | -18.514 | -7.356 | -10.324 | 4.859 |
| | (11.782) | (14.313) | (7.665) | (14.299) | (11.971) |
| ln response time | 13.160** | 10.022 | 7.243 | 0.261 | 5.712 |
| | (6.022) | (7.568) | (4.906) | (6.726) | (6.226) |
| Constant | 83.472*** | 47.444** | 123.227*** | 174.597*** | 187.142*** |
| | (25.054) | (23.652) | (17.533) | (29.521) | (23.358) |
| Observations | 167 | 110 | 174 | 111 | 146 |
| R-squared | 0.128 | 0.121 | 0.043 | 0.076 | 0.041 |

very low likelihood (‘Exceptionally unlikely’ or 1%), low likelihood (‘Unlikely’ or 25%), medium likelihood (‘About as likely as not’ or 50%), high likelihood (‘Likely’ or 75%), very high likelihood (‘Virtually certain’ or 99%)

Robust standard errors in parentheses

*** p<0.01, ** p<0.05, * p<0.1

The results of the regression model remain largely unchanged when including all observations without accounting for translation accuracy (see Appendix E). To illustrate this robustness, Figure 3 displays a forest plot comparing the effects of verbal communication on reservation prices across all likelihood levels for both matched and unmatched models. The figure shows that effect sizes are consistent regardless of matching on translation accuracy. In particular, the effects of verbal communication at medium and high likelihoods remain negative and statistically significant at the 10% level independent of the matching process. Therefore, we can confirm hypothesis 1 and 3 for the medium and high likelihood level, as the verbal group selected significantly lower reservation prices than the numerical group. Moreover, the aforementioned negative effect at a high likelihood level leads to less rational decisions, as reservation prices deviate further from the expected value of the lottery. Therefore, hypotheses 2 and 4 can be confirmed for a high likelihood level.

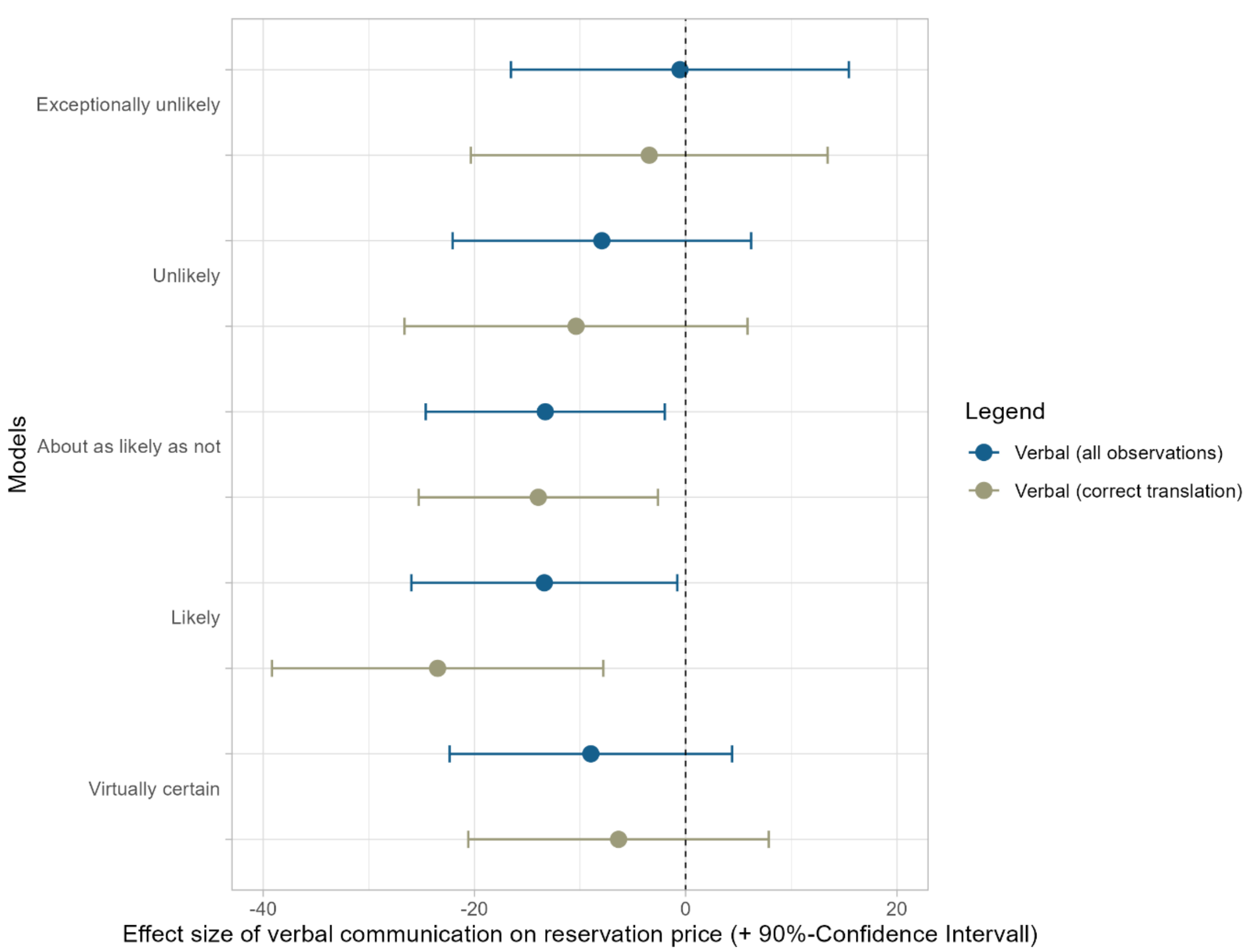

Figure 3. Effect of verbal communication

Knowing that the effect of verbal communication is not driven by the translation, we want to further explore how the effect emerges. Figure 4 incorporates five density plots in a 2x3 grid. Each plot displays three density functions depicting the distribution of reservation prices selected based on numerical and verbal communication both before and after matching. The plots differ in the likelihood level that was communicated. Here, we focus on medium likelihood ('about as likely as not') and high likelihood ('likely') as the effects of verbal communication are significant at these levels.

For a medium likelihood, the distribution of reservation prices appears very similar between numerical and verbal communication. Both conditions show a distinct peak at the most common reservation price of 100 pence, which also aligns with the expected value of the lottery. However, differences become more pronounced at the extremes: participants in the verbal condition (both before and after matching) are less likely to select very high reservation prices (near 200) and more likely to select very low ones (near 0) compared to those in the numerical condition.

For a high likelihood, very low reservation prices are uncommon for either form of communication, indicating that most participants recognize the high chance of winning. However, reservation prices begin to rise earlier under verbal communication, reaching a momentary peak at 100 pence. In contrast, the numerical condition shows a more gradual increase, culminating in a distinct peak at 150 pence, which aligns with the expected value of the lottery. In the verbal condition, the distribution is more widespread, lacking a distinct concentration around 150 pence. A similar pattern can be observed for reservation prices set for a very high likelihood ('virtually certain' or 99%) with an expected value of the lottery at 198 pence, suggesting that verbal framing may introduce greater variability and less consensus in decision-making, specifically for higher likelihoods of winning. Notably, the greater variability in reservation prices under verbal communication is not caused by differences in the translation of verbal phrases itself, as the distribution of reservation prices remains very similar when reducing the variability in translations through the matching process.

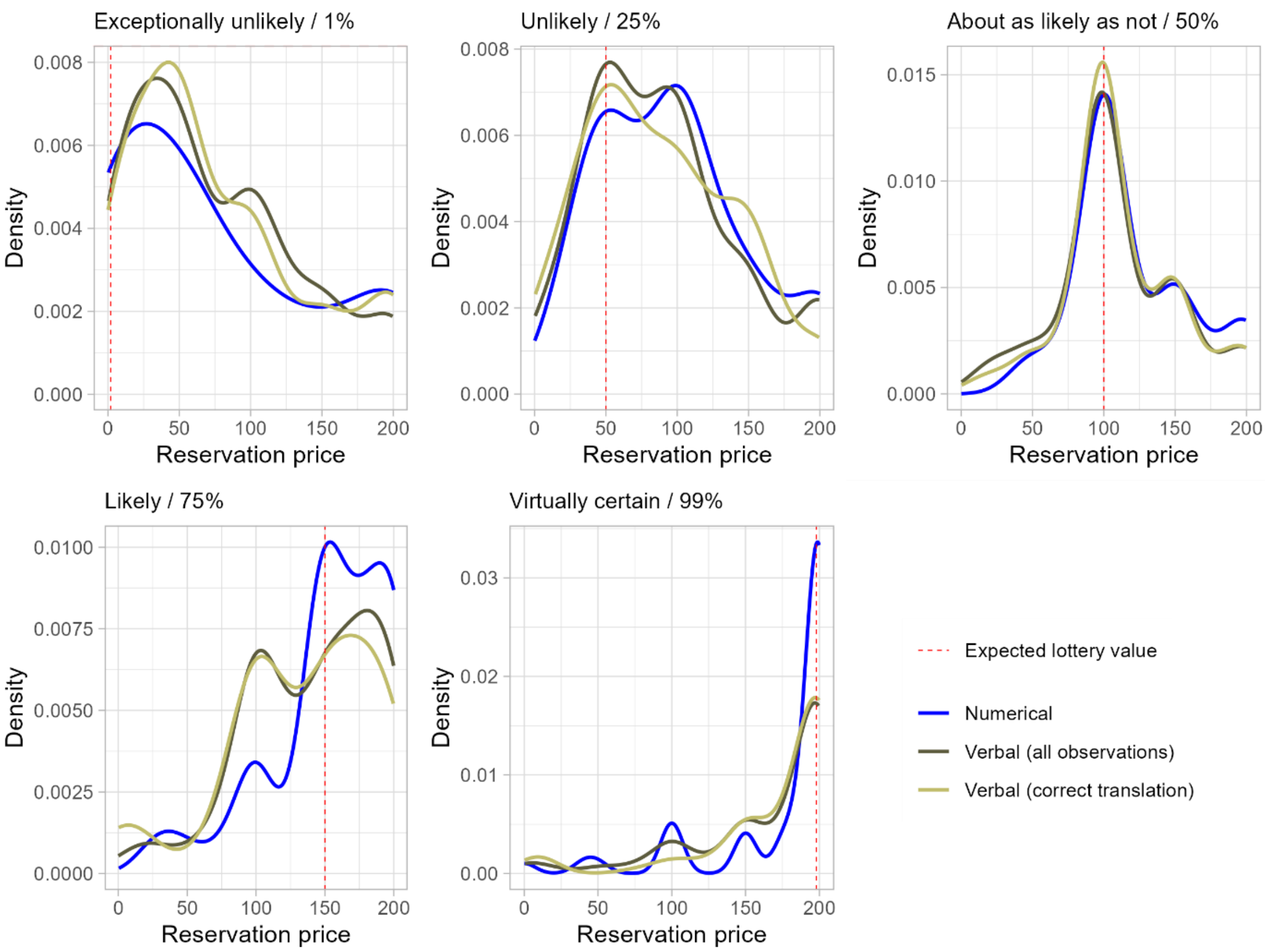

Figure 4. Density functions of reservation prices for numerical and verbal communication

Lastly, we are interested in the effect of the remaining covariates. Here, only a few significant effects can be observed in the regression models depicted in Table 4. Most notably, the BNT score, as an indicator of cognitive ability and numeracy skills, has a significant negative effect on reservation prices for lotteries with a very low and low likelihood of winning. Specifically, participants in the two highest quartiles of BNT scores select significantly lower reservation prices than participants in the lowest quartile. At low likelihood levels participants selected reservation prices that on average exceeded the expected value of the lottery by far. The negative effect of a high BNT score leads to a smaller deviation from the expected value.

Therefore, we conclude that high cognitive abilities can lead to more rational decisions at specific likelihood levels.

## 5. DISCUSSION

In this paper, we explore how the communication form of uncertainty (verbal vs. numerical) influences its perception and shapes subsequent decision-making. Based on the existing research on decision-making under uncertainty we know that individuals frequently choose irrational responses to uncertainty and that the decision-making process is susceptible to biases (e.g., loss aversion, see Kahneman & Tversky, 1979). Moreover, individuals have difficulties assessing uncertainties and respond negatively to ambiguity (ambiguity aversion, see Ellsberg, 1961). We find that verbal uncertainty expressions influence decision-making in ways that are consistent with this phenomenon: they lead to less optimistic evaluations and less rational decisions. Importantly, this effect is not merely a result of miscommunication caused by unintended translations of verbal phrases. Even when individuals correctly infer the intended probability value, the additional uncertainty inherent in the verbal format – its ambiguity – influences subsequent decisions.

Ambiguity refers to second-order uncertainty – uncertainty not just about outcomes, but about the probabilities themselves. Verbal phrases clearly exhibit this feature as documented in previous research. Their numerical translations differ considerably, both within and across studies. For instance, a meta study by Theil (2002) showed that the mean translation of the phrase 'likely' varies substantially, ranging between 63% to 77% across different studies. In our experiment, the phrase was translated with 69.47% on average, aligning well within this range. However, individual responses spanned a much broader range with standard deviations hovering around 12 percentage points across all five verbal probabilities adopted from IPCC reports.

While the translation of verbal probabilities has been extensively studied, previous work has been largely detached from research on decision-making under uncertainty. Recent research has started exploring the connection between verbal uncertainty and risk attitudes, indicating differences in risk aversion between verbal and numerical uncertainty (see Milczarski et al., 2025). We expand on this by exploring differences in the communication format under the theoretical framework of ambiguity aversion. Our results show that individuals select a significantly lower reservation price for an uncertain option when uncertainty is communicated verbally rather than numerically at a medium and high likelihood level. Notably, this leads to less rational behavior at high likelihoods, as on average reservation prices deviate further from the expected value of the lottery. Crucially, these effects persist even when accounting for translation accuracy, indicating that our results are not caused by potential misalignment between intended and perceived probability in the verbal group but rather by verbal communication itself.

Our results are consistent with previous evidence on ambiguity aversion, where participants selected higher reservation prices for risky options than for ambiguous ones (e.g., see Halevy, 2007; Borghans et al., 2009). We provide novelty by operationalizing ambiguity through verbal probabilities rather than numerical ranges. Additionally, our results further reinforce the idea that ambiguity aversion is not universal and instead align with findings by Kocher et al. (2018) showing that ambiguity aversion may only hold at medium to high likelihoods.

While ambiguity provides a compelling explanation for the observed differences in reservation prices at medium and high likelihoods, it may not fully account for behavior at low likelihoods. Instead, directionality as another key characteristic of verbal phrases needs to be considered. Directionality refers to the linguistic framing of uncertainty expressions that can guide decision-making independently of perceived probability values or ambiguity (Teigen & Brun, 1999). More specifically, phrases with positive directionality (e.g., possible) are encouraging as they highlight the occurrence of an uncertain event whereas phrases with a negative directionality (e.g., doubtful) emphasize the non-occurrence and advise to be careful (Teigen & Brun, 1995, 2003). Notably, differences in directionality between verbal and numerical communication typically only occur at low likelihood levels: while verbal expressions often adopt a negative framing, numerical probabilities are considered unidirectional with a positive bias, as their directionality remains positive even for low likelihood levels. Teigen and Brun (1995; 2000) demonstrated this through multiple experiments in which participants often judge even low numerical probabilities as affirmative and associate them with positive outcomes.

In our experiment, individuals exhibited risk-seeking behavior at low likelihood levels for both communication modes. For numerical probabilities such behavior aligns with established findings in prospect theory as individuals tend to overestimate low probability values (see Kahneman & Tversky, 1979). In contrast, risk-seeking decisions for low likelihood phrases (such as unlikely) are surprising, as their negative directionality has been shown to elicit less optimistic decision-making compared to numerical probabilities (e.g., see Doyle et al., 2014; Collins & Mandel, 2019; Honda et al., 2023; Collins et al., 2024). We do not dismiss the potential influence of directionality. Instead, there could be an interaction between directionality and ambiguity that we cannot disentangle within the scope of our experiment. While ambiguity aversion is a well-established behavioral tendency, Kocher et al. (2018) show that individuals may become ambiguity seeking at low likelihoods. Such preferences would lead to higher reservation prices and thereby counteract the caution induced by negative directionality.

Our paper offers both practical and theoretical implications for communication of uncertainty as well as its impact on subsequent decisions. As highlighted in previous research, senders of messages should be aware that receivers may systematically mistranslate verbal probabilities (e.g., see Brun & Teigen, 1988; Budescu et al., 2009, 2012). Moreover, we show that even in absence of mistranslations, verbal probabilities interfere with decision-making, leading to lower evaluations of uncertain options at medium and high likelihoods, as well as less rational responses at high likelihoods. These findings are relevant for anyone communicating uncertain events, especially managers and policymakers who need to convey uncertainties in a clear and impactful way. In an organizational setting managers frequently communicate uncertainties to employees, stakeholders and higher-level executives. While managers always have the option to communicate uncertainties numerically in principle, they often lack objective numerical probabilities as a foundation. In many cases, their assessments of uncertainty are based on subjective judgments rather than precise statistical models, which makes verbal probability phrases a more intuitive or practical choice. Nonetheless, verbal probability can unintentionally influence employees' decision-making processes. Organizations should reconsider when and how they use verbal probabilities, particularly in situations where neutrality in uncertainty communication is crucial.

Moreover, with the rise of machine learning models in organizational settings, numerical probability estimates are becoming more widely available. This has sparked discussions on how AI systems should communicate uncertainty to human decision-makers (see Papenkordt et al., 2023). Based on our findings, we recommend prioritizing precise numerical probabilities over verbal probability phrases to ensure that decision-making is guided by the actual likelihood level rather than the framing of uncertainty.

We also contribute to current research on ambiguity aversion, which has seen a spike in interest over the last two decades (see Bühren et al., 2023). In this time, many expansions of Elsberg's (1961) original design were published, yet the source of ambiguity remained limited to numerical ranges or missing probabilities. Both Ellsberg (2011) and Bühren et al. (2023) question the applicability of ambiguity aversion beyond theoretical models. We believe that this gap in practical relevance can be partially attributed to the narrow focus on numerical ambiguity even though verbal probabilities are the preferred form of communicating uncertainty in spoken language (see Erev & Cohen, 1990; Wallsten et al., 1993; Juanchich & Sirota, 2020a, 2020b; Rosen et al., 2021). By considering verbal probabilities as a source of ambiguity, we expand on previous research and show that ambiguity aversion may not be limited to numerical forms of ambiguity. To enhance the practical relevance of ambiguity research, we suggest that future research continues to explore the role of verbal probabilities.

Lastly, our experiment leaves room for further future research. We see an urgent need to externally validate our results. We used a laboratory experiment, so evidence from the field would be needed for validation. In particular, we implemented a rather low-stake setting and the effect of varying stakes (low and high) as well as decisions for oneself or a larger group (or generally third parties) would be needed (see Polman, 2012). Also, we applied a neutral and simplified decision-making context which does not represent the potential complexity of real-world scenarios. While it has been shown that contextual information paired with prior beliefs can influence the translation of verbal phrases (see Brun & Teigen, 1988; Budescu et al., 2012), further research should examine whether this influence extends to subsequent decision-making.

**Competing Interests**

The authors have no competing interests.

**Financial Support Statement**

This research did not receive any specific grant from funding agencies in the public, commercial, or not-for-profit sectors.

## Appendix A

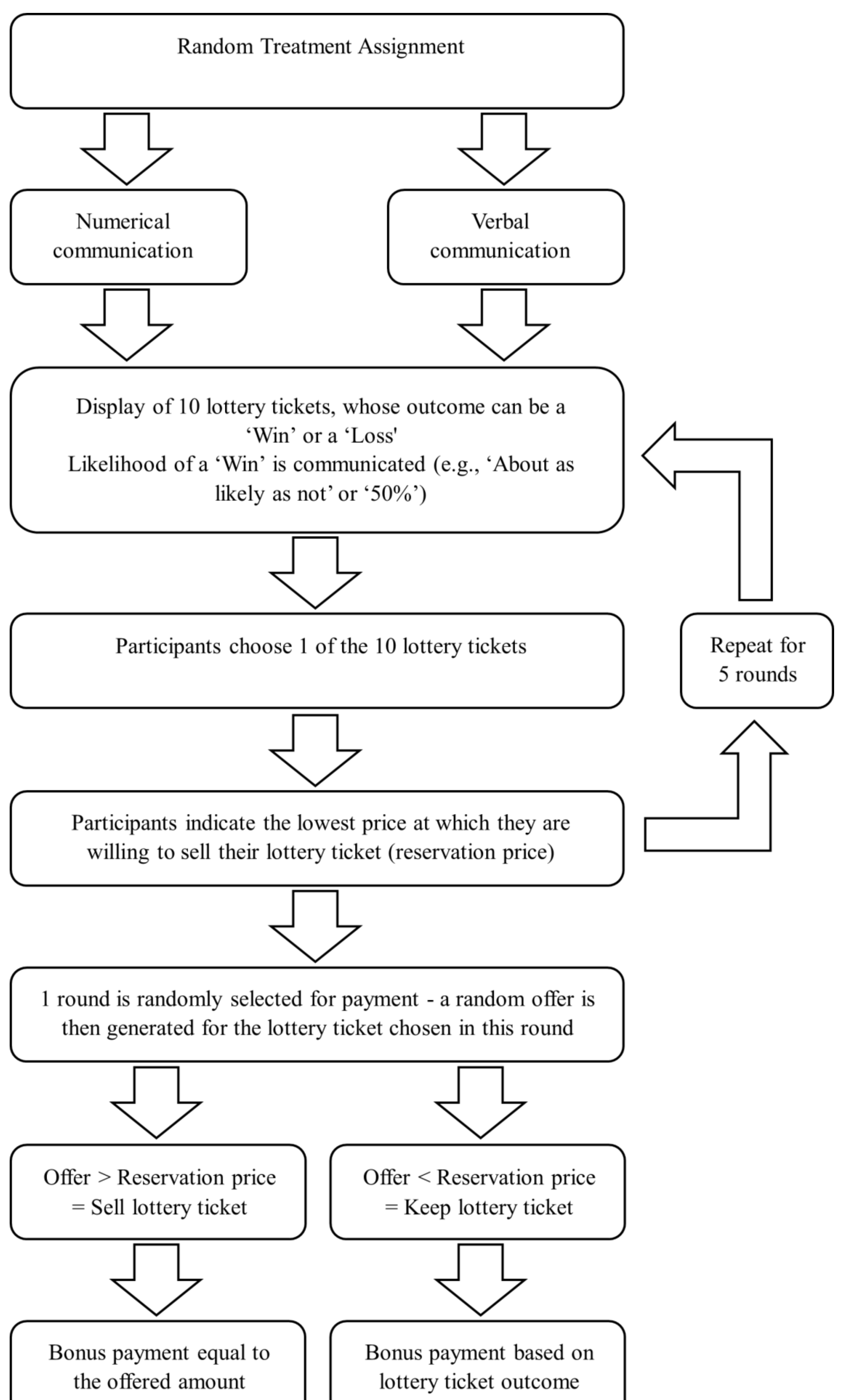

Figure A.1. Experimental Procedure

## Appendix B

The questions of the BNT designed by Cokely et al. (2012) are as follows:

1) Out of 1.000 people in a small town 500 are members of a choir. Out of these 500 members in the choir 100 are men. Out of the 500 inhabitants that are not in the choir 300 are men. What is the probability that a randomly drawn man is a member of the choir?

Correct answer: 25

2a) Imagine we are throwing a five-sided die 50 times. On average, out of these 50 throws how many times would this five-sided die show an odd number (1, 3 or 5)?

Correct answer: 30

2b) Imagine we are throwing a loaded die (6 sides). The probability that the die shows a 6 is twice as high as the probability of each of the other numbers. On average, out of these 70 throws how many times would the die show the number 6?

Correct answer: 20

3) In a forest 20% of mushrooms are red, 50% brown and 30% white. A red mushroom is poisonous with a probability of 20%. A mushroom that is not red is poisonous with a probability of 5%. What is the probability that a poisonous mushroom in the forest is red?

Correct answer: 50

Participants receive 2-3 questions based on a dynamic design:

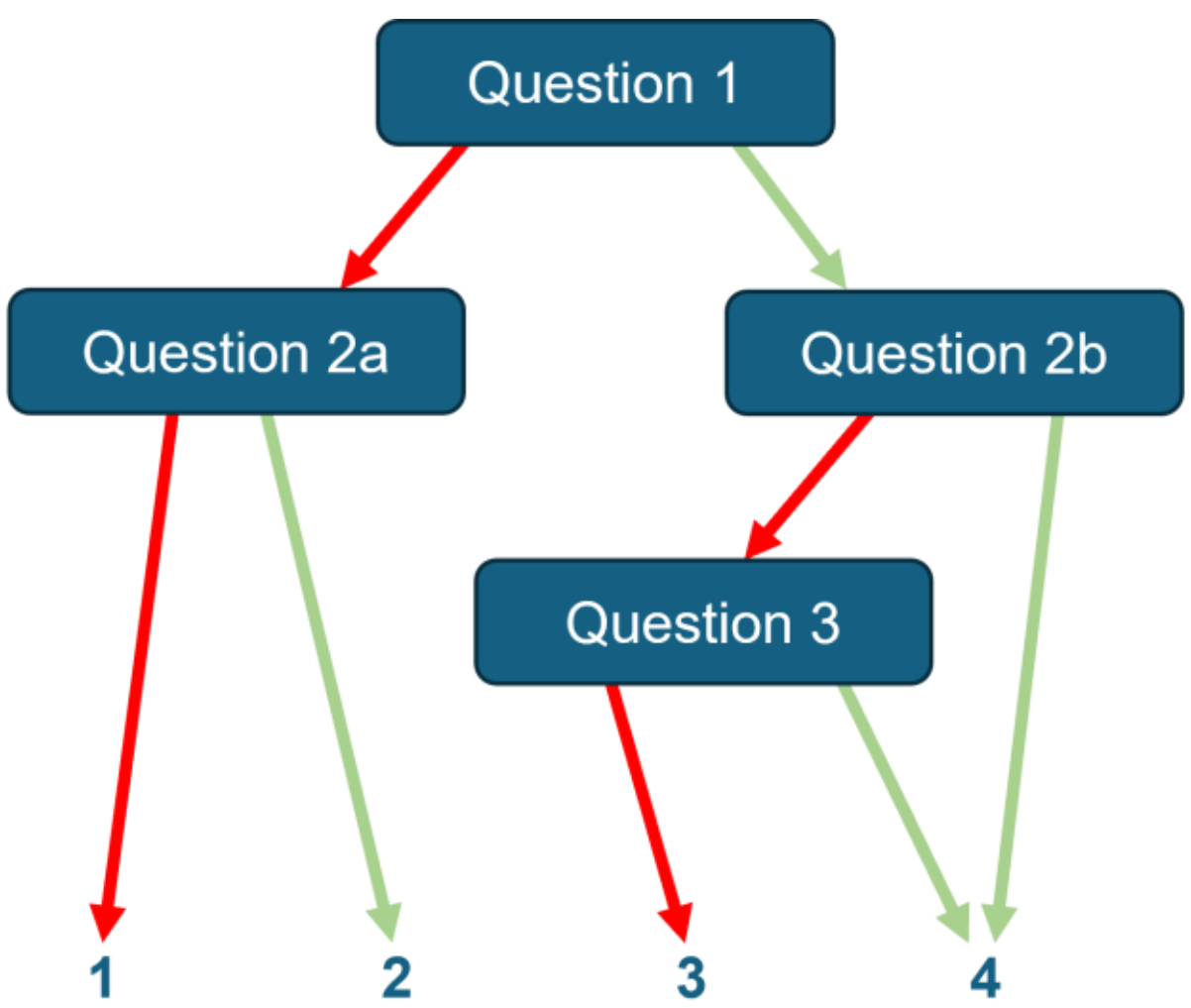

Figure B.1. BNT dynamic questionnaire design

**Appendix C**

Table C.1. Regression models for the numerical translation of verbal phrases

| Explanatory variables | (1) Very low likelihood | (2) Low likelihood | (3) Medium likelihood | (4) High likelihood | (5) Very high likelihood |
|---|---|---|---|---|---|
| Male | 2.501 | 1.043 | -1.155 | -0.114 | -1.748 |
| (0=female) | (2.736) | (2.115) | (1.992) | (1.943) | (2.004) |
| Age | -0.210* | -0.165* | -0.121 | 0.184** | 0.131** |
| | (0.114) | (0.099) | (0.146) | (0.078) | (0.060) |
| BNT (1st quartile – ref.) | | | | | |
| 2nd quartile | -0.611 | 0.058 | -4.868* | 3.396 | 2.844 |
| | (4.236) | (2.847) | (2.919) | (2.352) | (2.361) |
| 3rd quartile | -3.153 | 1.581 | -5.034 | 4.477* | 3.872 |
| | (3.548) | (2.821) | (4.504) | (2.626) | (2.814) |
| 4th quartile | -0.709 | -0.168 | -1.759 | -0.052 | 2.627 |
| | (3.193) | (2.617) | (1.784) | (2.408) | (2.503) |
| ln response time | -1.906 | -0.845 | -1.383 | -0.569 | -2.756 |
| | (1.615) | (2.299) | (1.706) | (1.814) | (1.847) |
| Constant | 21.024*** | 33.549*** | 57.016*** | 63.173*** | 91.904*** |
| | (5.619) | (5.738) | (6.285) | (4.677) | (4.063) |
| Observations | 149 | 149 | 149 | 149 | 149 |
| R-squared | 0.026 | 0.024 | 0.049 | 0.049 | 0.053 |

very low likelihood (‘Exceptionally unlikely’ or 1%), low likelihood (‘Unlikely’ or 25%), medium likelihood (‘About as likely as not’ or 50%), high likelihood (‘Likely’ or 75%), very high likelihood (‘Virtually certain’ or 99%)

Robust standard errors in parentheses

*** $p<0.01$, ** $p<0.05$, * $p<0.1$

## Appendix D

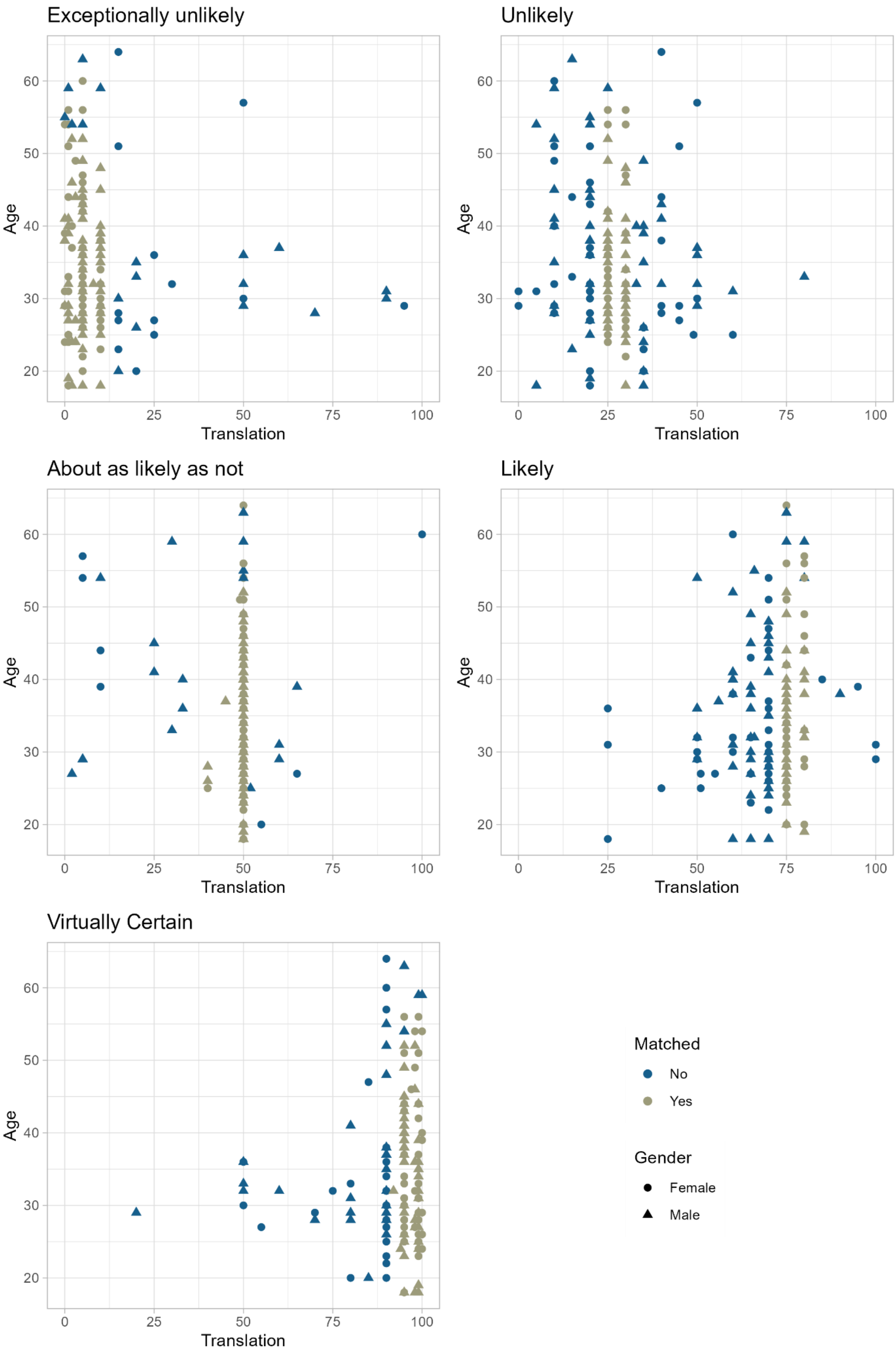

Figure D.1. Matched observations

**Appendix E**

Table E.1. Regression models for reservation prices (all observations)

| Explanatory variables | (1) Very low likelihood | (2) Low likelihood | (3) Medium likelihood | (4) High likelihood | (5) Very high likelihood |
|---|---|---|---|---|---|
| Verbal | -0.542 | -7.935 | -13.288** | -13.382* | -8.977 |
| (0=numerical) | (10.438) | (8.345) | (6.660) | (7.260) | (7.821) |
| Male | 0.892 | -0.257 | -8.121 | -11.670* | -11.094 |
| (0=female) | (8.401) | (7.383) | (5.983) | (6.760) | (7.110) |
| Age | -0.445 | -0.158 | -0.299 | -0.487 | -0.025 |
| | (0.384) | (0.362) | (0.300) | (0.329) | (0.339) |
| BNT ($1^{st}$ quartile – ref.) | | | | | |
| $2^{nd}$ quartile | -7.344 | -10.526 | -3.410 | 7.159 | 9.988 |
| | (12.637) | (10.471) | (8.768) | (9.552) | (10.513) |
| $3^{rd}$ quartile | -44.585*** | -23.209* | -20.577* | -7.674 | 26.407*** |
| | (14.171) | (12.348) | (11.435) | (11.848) | (9.085) |
| $4^{th}$ quartile | -29.303*** | -20.020** | -11.339 | 0.908 | 13.867* |
| | (9.752) | (8.762) | (6.916) | (8.132) | (8.350) |
| ln response time | 12.797*** | 7.714 | 5.862 | 1.251 | 2.719 |
| | (4.883) | (4.746) | (4.341) | (3.930) | (4.599) |
| Constant | 78.114*** | 97.740*** | 131.262*** | 173.545*** | 165.086*** |
| | (20.587) | (18.306) | (15.173) | (16.214) | (17.752) |
| Observations | 200 | 200 | 200 | 200 | 200 |
| R-squared | 0.086 | 0.046 | 0.064 | 0.048 | 0.042 |

very low likelihood (‘Exceptionally unlikely’ or 1%), low likelihood (‘Unlikely’ or 25%), medium likelihood (‘About as likely as not’ or 50%), high likelihood (‘Likely’ or 75%), very high likelihood (‘Virtually certain’ or 99%)

Robust standard errors in parentheses

*** p<0.01, ** p<0.05, * p<0.1